\documentstyle[11pt,newpasp,twoside,epsf,psfig]{article}

\def\etal{et al.}
\def\hii{H{\sc ii}}

\def\msun{M$_{\odot}$}

\def\halpha{\ifmmode {\rm H{\alpha}} \else $\rm H{\alpha}$\fi}
\def\hbeta{\ifmmode {\rm H{\beta}} \else $\rm H{\beta}$\fi}
\def\heii{\ifmmode {{\rm He{\sc ii}} \lambda 4686} \else {He\,{\sc ii} $\lambda$4686}\fi}

\def\civ{C\,{\sc iv} $\lambda$5808}

\begin{document}

\title{Super star clusters as probes of 
massive star evolution and the IMF at extreme metallicities} 

\author{Daniel Schaerer} 
\affil{Observatoire Midi-Pyr\'en\'ees, 
       14, Av. E. Belin, F-31400, Toulouse, France
      (schaerer@obs-mip.fr)}
\author{Natalia G. Guseva and Yuri I. Izotov}
\affil{Main Astronomical Observatory
                 of National Academy of Sciences of Ukraine,
                 Goloseevo, 03680 Kiev-127, Ukraine
      (guseva@mao.kiev.ua, izotov@mao.kiev.ua)}


\begin{abstract}
Young super star clusters and young compact massive star forming regions
can provide useful information on their burst properties (age, burst duration, SFR),
the upper end of the IMF and yield new constraints on the evolution of massive
stars.
Through the study of their stellar populations we can in particular extend
our knowledge on massive stars to extreme metallicities unavailable for such stars
in the Local Group.

Here we summarise the main results from recent studies on two metallicity extremes:
Wolf-Rayet and O star populations in very metal-poor BCD, and metal-rich
compact nuclear SF regions.
\end{abstract}

\section{Introduction}
Two ``modes'' of star formation are observed in (optically or UV selected) 
starburst galaxies (e.g.\ Meurer \etal\ 1995):
a young unresolved population responsible for emission of diffuse UV light
   (Meurer \etal\ 1995, also Calzetti these proceedings), and
compact stellar clusters, losely termed super star clusters (SSCs) hereafter.
SSCs have been the focus of numerous recent studies related in particular to
the possibility that these clusters may represent the progenitors of globular
clusters (cf. Fritze von Alvensleben, Miller, these proceedings).
A different aspect is emphasized in the present work.
We use spectroscopic observations of young star forming (SF) regions
to determine their massive star content with the aim of providing
constraints on stellar evolution models and the upper end of the IMF.

SSCs and similar compact young SF regions have the following properties:
{\em a)} Numerous such objects are known.
{\em b)} They represent clusters rich enough ($\sim$ 10$^{2-4}$ O stars) such
   that the IMF can be well populated and stochastical effects 
   (cf.\ Lan\c{c}on these proceedings) are negligible.
{\em c)} A priori the clusters cover a wide range of metallicities, and
{\em d)} consist likely of a fairly coeval population.
Given these properties, SSCs resemble ``normal'' Galactic of Local Group
clusters which represent fundamental test-cases for stellar evolution.
The only disadvantage is that their stellar content can only be studied
through observations of their integrated light.
On the other hand b) and c) represent important advantages for studies focussed 
on massive stars over using ``local'' clusters.
{\em This shows that young SSCs provide ideal samples for studies of massive star
evolution in different environments, such as e.g.\ extreme metallicities 
largely inaccessible in Local Group objects.}

After a brief introduction on the type of objects used here (Wolf-Rayet rich
SF region) we will summarise recent work along these lines.

\section{Wolf-Rayet galaxies and clusters}
We will concentrate on the so-called Wolf-Rayet (WR) galaxies (cf.\
Schaerer \etal\ 1999b for the latest catalogue), which are objects where 
broad stellar emission lines (called ``WR bumps'', mostly at \heii\ and \civ) 
in the integrated spectrum testify to the presence of WR stars.
For the study of massive star populations these objects are ideal since 
WR stars are the descendents of the most massive stars in a short-lived phase
($M_{\rm ini} \ga 25$ \msun, $t_{\rm WR} \sim 10^{5-6}$ yr).
Their detection is also a good age indicator for young systems ($t \la 10$ Myr), 
and allows good measure of the burst duration and the best direct probe of the 
upper end of the IMF.
An overview of studies on WR populations in starburst regions can be found
in the reviews of Schaerer (1998, 1999). 

In the context of the present workshop it is important to note that the
objects broadly referred to as WR ``galaxies'' are found among a large variety 
of objects including BCD, massive spirals, IRAS galaxies, Seyfert 2, and LINERs
(see Schaerer \etal\ 1999b). The ``WR rich'' regions contained in the
spectroscopic observations will thus in general cover quite a large scale
of sizes, different morphologies etc. In the case of blue compact dwarfs
(BCD), one is however mostly dealing with one or few individual compact 
regions or SSC dominating the observed light. Although this statement 
cannot, with few exceptions, be quantified so far for the objects studied
below (but see e.g.\ Conti \& Vacca 1994) we will mostly assume that the 
spectroscopic observations correspond closely to light from one young
compact SF region or SSC.

\section{Studies of Wolf-Rayet populations in metal-poor environments}
The spectroscopic sample of dwarf galaxies from Izotov, Thuan and collaborators,
obtained for the main purpose of determining the primordial He abundance
and other abundance studies, has proven to be very useful for 
analysis of massive star populations especially at very low metallicities.
Indeed, $\sim$ 20 WR rich regions are found in this sample at 
metallicities below the SMC ($12 + \log$ O/H $\la$ 8.1) extending to
I Zw 18 with $\sim$ 1/50 solar metallicity. No {\em bona fide} massive star
of such low a metallicity is known in the Local Group!

\begin{figure}[htb]
\centerline{\psfig{file=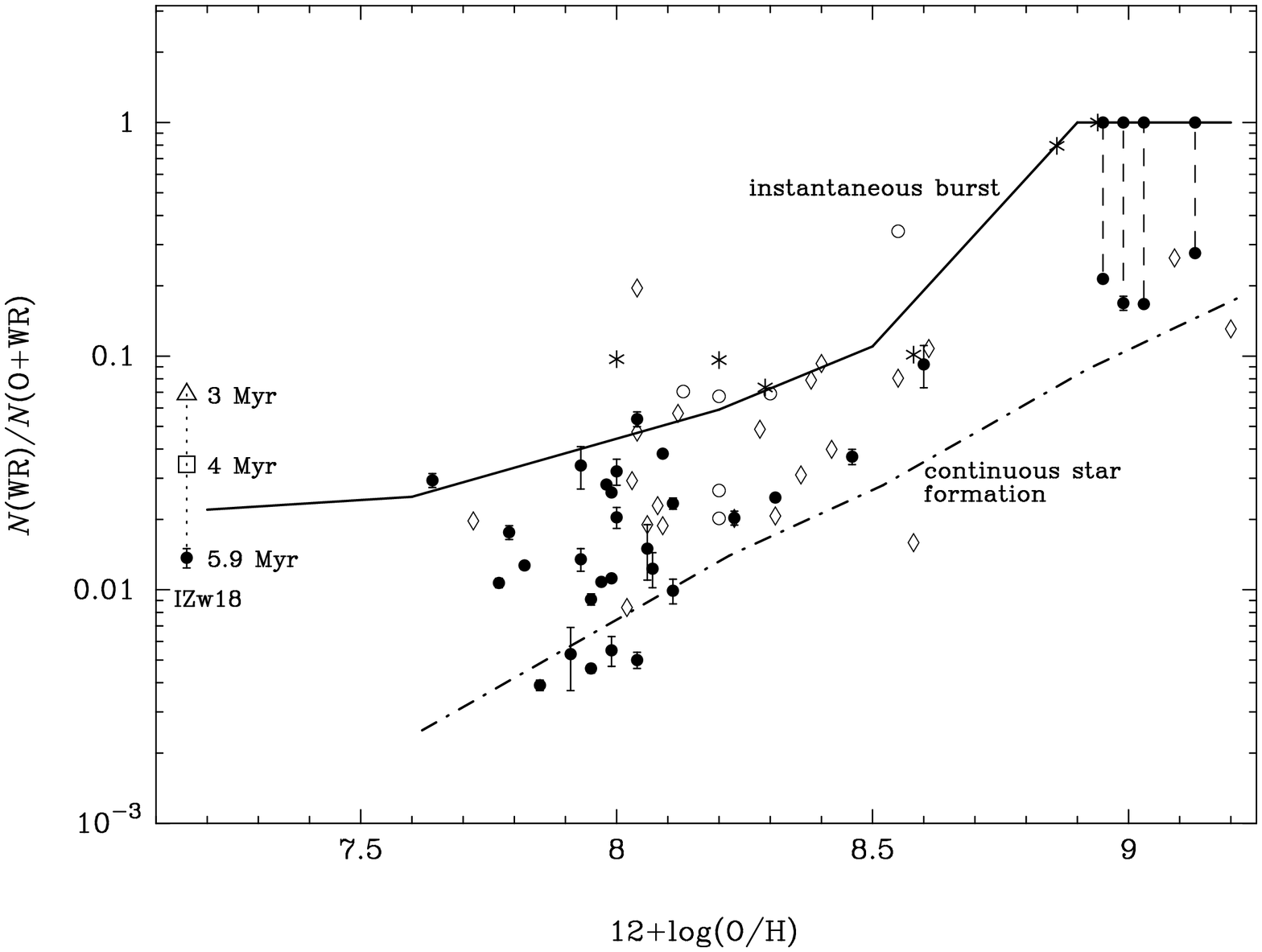,angle=270,width=8cm}
            \psfig{file=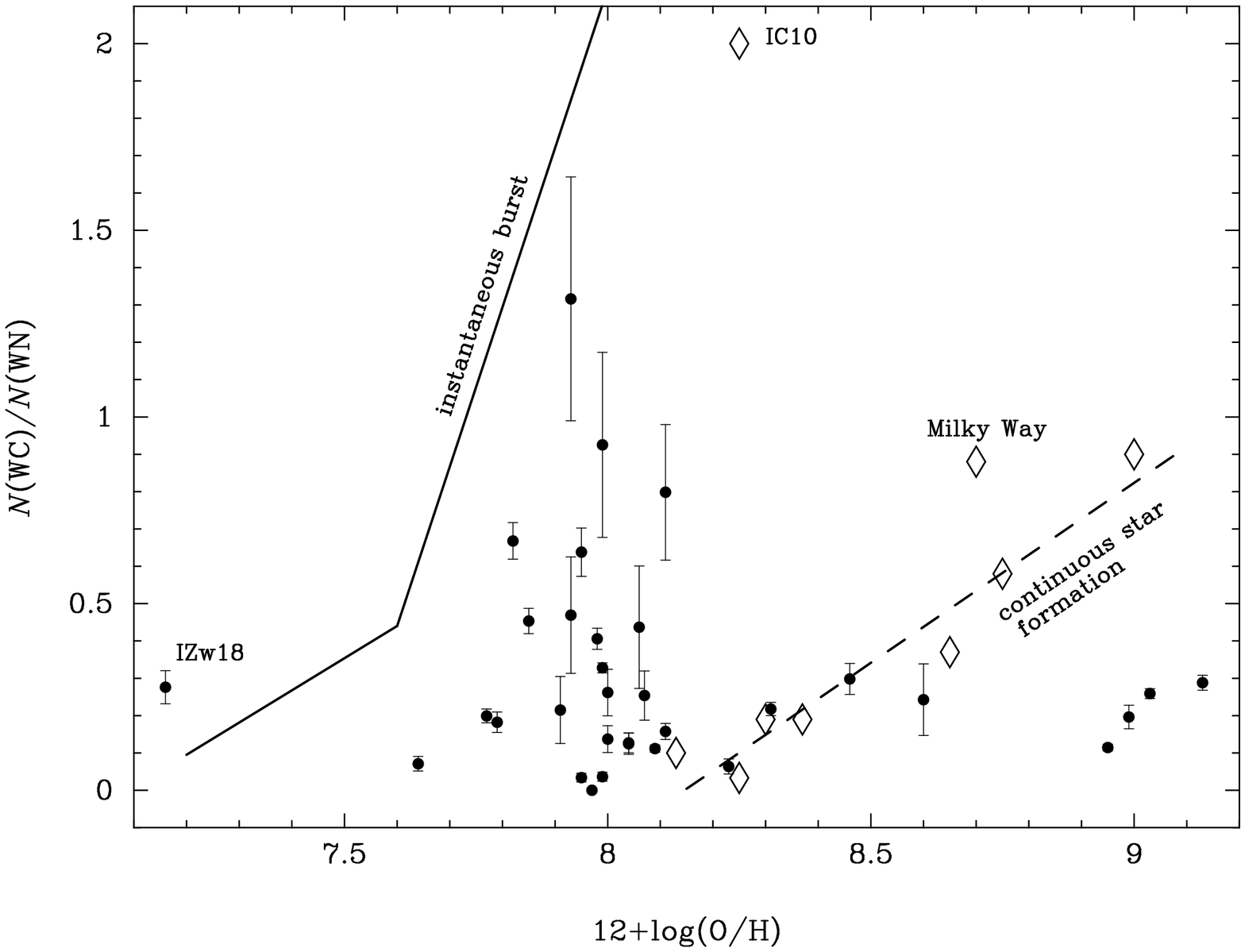,angle=270,width=8cm}}

  \caption{{\em Left:} Number ratio WR/(WR+O) as a function of metallicity
for objects from GIT99 (filled circles), Vacca \& Conti (1992, diamonds),
Kunth \& Joubert (1985, asterisks), and SCK99 (open circles). Maximum values
predicted from the SV98 models for a Salpeter IMF ($M_{\rm up}=120$ \msun)
are shown for instantaneous bursts (upper line) and continuous SF (lower).
{\em Right:} Number ratio WC/WN as a function of metallicity from GIT99
(filled circles). Diamonds illustrate the observed WC/WN ratios in Local
Group objects though to represent regions at ``SF equilibrium'' (constant SF).
Model predictions as above.
Figs.\ taken from GIT99.}
\end{figure}

The analysis of the WR and O star content in these objects has been
presented by Guseva et al.\ (1999, hereafter GIT99).
Some of their main results are summarised in Fig.\ 1, which shows
(left panel) the derived WR/(WR+O) number ratio as a function of metallicity
from their objects and observations of Kunth \& Joubert (1985), Vacca \& Conti
(1992), and Schaerer et al.\ (1999a, hereafter SCK99).
The left Fig.\ considerably extends the previous samples (cf.\ Vacca \& Conti 1992,
Meynet 1995). The trend of increasing WR/O with metallicity is well understood
(Arnault et al.\ 1989)
The comparison with appropriate evolutionary synthesis models
(Schaerer \& Vacca 1998, SV98; shown as solid lines) calculated for a ``standard''
Salpeter IMF with $M_{\rm up}=120$ \msun\ and using the high mass loss Geneva tracks 
shows a good agreement.
This and more direct comparisons of the observed WR features (see Schaerer 1996,
de Mello et al.\ 1998, SCK99, GIT99) indicate that the bulk of the observations 
are compatible with short (``instantaneous'') bursts with a Salpeter IMF extending
to large masses. The short burst durations\footnote{See Meurer (these proceedings)
for a discussion of SF durations.} derived by SCK99 for the metal-poor objects
are also in agreement with the study of Mas-Hesse \& Kunth (1999).

Of particular interest for evolutionary models is the relative number of WR stars
of the WN (exhibiting H-burning products on their surface) and WC subtypes
(He-burning products). The relative lifetimes vary strongly with initial mass
and metallicity and are sensitive to various mass loss prescriptions
and mixing scenarios currently not well known (see Maeder \& Meynet 1994, 
Meynet these proceedings).
The recent high S/N spectra of SCK99 and GIT99 have now allowed to establish
number ratios of WC/WN stars in a fair number of WR rich regions. The determinations
of GIT99 are shown in Fig.\ 1 (right panel); similar values are derived by
SCK99. The comparison with synthesis models shows a reasonable agreement.
To reproduce sufficiently large WC/WN ratios the use of the stellar tracks
based on the high mass loss prescription are, however, required as shown
by de Mello et al.\ (1998) and SCK99. 

It is understood that part of the ``requirement'' for the high mass loss (cf.\ Schaerer 1998) 
may be compensated by additional 
mixing processes leading to a similar prolongation of the WR phase (cf.\ Meynet, these 
proceedings). 
In any case in addition to  the well known stellar census in the Local Group 
(cf.\ Maeder \& Meynet 1994 and references therein) the present new 
data from integrated populations place important constraints on the evolutionary
models which have to be matched by successful stellar models.
Especially the new studies extend the range of available metallicities to very 
low $Z$, well beyond the SMC.

\section{Massive stars and the IMF in metal-rich starbursts}
A small sample of metal-rich (O/H $\ga$ solar) starbursts (4 objects from GIT99
and Mrk 309) have recently been analysed in detail by Schaerer et al.\ (2000).
In this case the observations (high S/N, intermediate resolution optical 
spectroscopy) correspond to compact nuclear SF regions.
Despite this complication we use these objects as a first step to probe the 
upper end of the IMF at high-metallicity.
Subsequent studies of more isolated and simple, cluster-like objects
will be undertaken in the future.

\begin{figure}[htb]
\plottwo{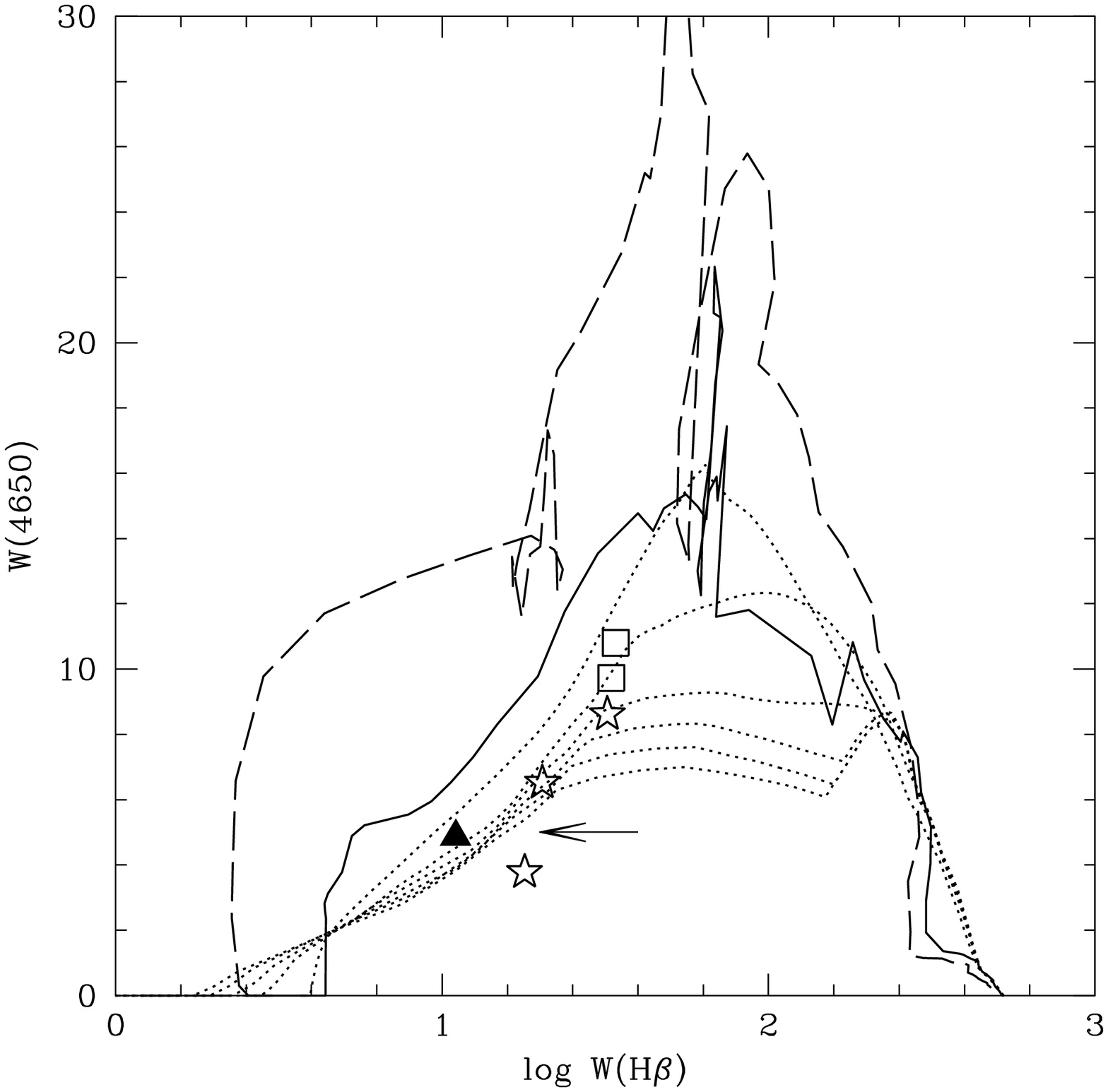}{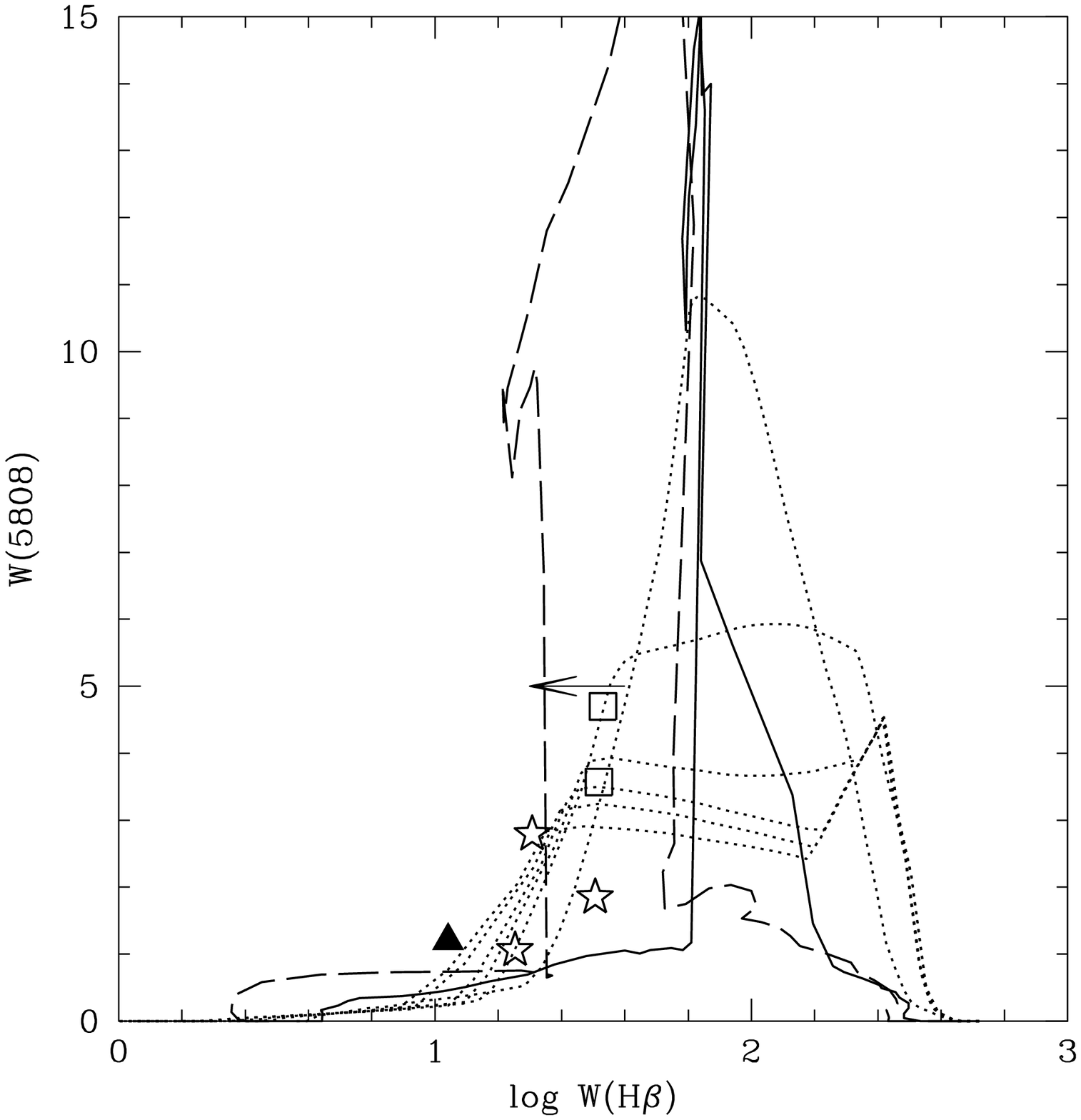}
  \caption{Observed and predicted EW of WR features (left: blue bump, right: \civ).
Model predictions are shown for instantaneous bursts at Z=0.02 (solid line)
and Z=0.04 (dashed), and extended bursts at Z=0.02 (dotted; burst durations
$\Delta t=2,4,6,8,10,12$ Myr).
All models assume a Salpeter IMF with M$_{\rm up}$=120 \msun.}
\end{figure}

\begin{figure}[htb]
\centerline{\psfig{file=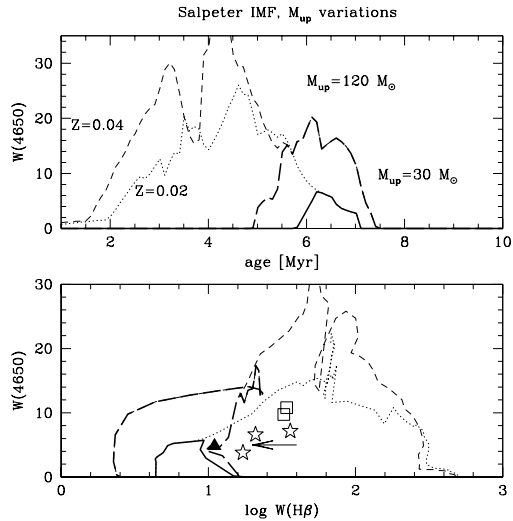,width=7cm}}
  \caption{Observed and predicted EW of the blue WR bump. 
The predictions show the dependence of the WR bump on the upper mass cut-off.
A conservative approach yields a {\em lower limit} of M$_{\rm up} \ga$ 30 \msun.}
\end{figure}

For our comparison with evolutionary synthesis models (cf.\ below)
we use following main observational constraints:
H$\beta$ and H$\alpha$ equivalent widths serving as age indicator,
H$\alpha$/H$\beta$ determining the gaseous extinction,
intensities and equivalent widths of the main WR features (4650 bump,
  \civ),
TiO bands at $\sim$ 6250 and 7200 \AA\ indicating the presence of 
  red supergiants from a population with ages $\ga$ 7--10 Myr, and
the overall SED provide an important constraint on the population
responsible for the continuum.

Model calculations have been done using the SV98 synthesis code.
The basic model parameters we consider are:
stellar metallicity,
IMF slope and upper mass cut-off,
star formation history (instantaneous or extended bursts, constant SF),
fraction of Lyc photons absorbed by the gas, 
stellar extinction (which may differ from gaseous).

\subsection{Results}
The comparison of the observed and predicted WR features is shown in 
Fig.\ 2.
The observations are well reproduced by Z=0.02 models with a ``standard IMF''
(Salpeter, M$_{\rm up}=$120 \msun) assuming extended burst durations 
of $\Delta t \sim$ 4--10 Myr.
The longer burst durations found here are physically plausible and
likely explained by the different nature of the metal-rich objects 
analysed here (larger nuclear regions vs.\ compact clusters in BCD,
cf.\ SCK99).
The corresponding ages of our objects, as indicated by W(H$\beta$), 
are between $\sim$ 7 and 15 Myr, also in agreement with the presence of 
the TiO bands.
A very good fit is also obtained to the overall SED. This requires, however,
an extinction of the stellar continuum which is less than that derived
from the gas. The differences are of similar amount that found by other
methods in other studies (e.g.\ Calzetti 1997, Mas-Hesse \& Kunth 1999).

In short, we conclude that all the given observational constraints can be 
well reproduced by models with a Salpeter IMF extending to high masses for 
a burst scenario with star formation extending over $\sim$ 4--10 Myr.
This solution is not unique. Therefore a variety of alternate models have 
been considered (cf.\ Schaerer et al.\ 2000).
Regarding the IMF we find that steeper (with slope $\sim$ Miller-Scalo) 
IMFs are very unlikely.

In view of several studies indicating a possible lack of massive stars
in metal-rich environments (e.g.\ Bresolin et al.\ 1999, Goldader et al.\ 1997)
we have used the present set of metal-rich WR galaxies to put
a {\em lower limit} on the value of M$_{\rm up}$ from the strength of the 
WR features.
As mentioned above our data is compatible with a large upper mass cut-off.
The real range of values our data is sensitive to is, however, limited;
intrinsically younger regions need to be sampled to probe the most massive
stars.
Adopting a conservative approach (cf.\ Fig.\ 3) we obtain M$_{\rm up} \ga$
30 \msun\ (for a Salpeter slope). 
We also find similar values (M$_{\rm up} \sim$ 35--50 \msun) from comparisons
of H$\beta$ equivalent width measurements in metal-rich \hii\ regions
(see Schaerer et al.\ 2000).

In contrast with previous studies of metal-rich starbursts and related objects
based on properties of the ionized gas
our work provides first constraints on the upper end of the IMF 
measured directly from stellar signatures. Future work on larger samples
and using detailed coupled stellar population and photoionisation 
models should be of great interest.

\begin{acknowledgements}
We thank the organisers, especially Ariane Lan\c{c}on, for this very
stimulating and pleasant workshop.
We acknowledge support to this collaboration trough INTAS grant 97-0033.
\end{acknowledgements}

\vspace{0.3cm}

{\bf Q. (Manfred Pakull):} Some years ago it was thought that WRs do not form
in very low Z environments because the massive stellar winds were too weak.
Is that still the state of the art in stellar evolution models ?
Or are the broad emission features due to a population of H-rich, very young
objects like in 30 Dor and NGC 3603 ?

{\bf A. (Daniel Schaerer:} Indeed the metallicity dependence of mass loss causes an important
decrease of the WR population toward low Z. Despite this evolutionary models
applying recent mass-loss prescriptions predict {\em some} WR stars at the lowest
metallicities (1/50 solar) corresponding to I Zw 18 (see de Mello et al.\ 1998),
quite in agreement with the observations.
The role of other formation channels (massive close binaries, rotation) at these 
low Z remains unexplored

In Schaerer et al.\ (1999, SCK99) we have explored the effect of such R136-like
WR stars during H-burning in addition to other WR and Of stars with broad emission
lines. Given their high initial mass their integrated contribution to the WR bump
should in most cases not be very important compared to ``normal'' WR stars.
An exception could be very young ($\la 2-3$ Myr) and strongly coveal populations.
Few such observational cases seem, however, known to date (see SCK99).

\end{document}